\documentclass{article}

    \PassOptionsToPackage{numbers,sort&compress}{natbib}

\usepackage[final]{neurips_2021}

\usepackage[utf8]{inputenc} 
\usepackage[T1]{fontenc}    
\usepackage{hyperref}       
\usepackage{url}            
\usepackage{booktabs}       
\usepackage{amsfonts}       
\usepackage{nicefrac}       
\usepackage{microtype}      
\usepackage{xcolor}         
\usepackage{amsmath}        
\usepackage{graphicx}
\title{Learning Large-Time-Step Molecular Dynamics with Graph Neural Networks}

\author{%
  Tianze Zheng\\
  Tsinghua University\\
  \And
  Weihao Gao\\
  ByteDance Inc.\\
  \AND
  Chong Wang\\
  ByteDance Inc.\\
  \texttt{chong.wang@bytedance.com}
}

\begin{document}

\maketitle

\begin{abstract}
  Molecular dynamics (MD) simulation predicts the trajectory of atoms by solving Newton's equation of motion with a numeric integrator. Due to physical constraints, the time step of the integrator need to be small to maintain sufficient precision. This limits the efficiency of simulation. To this end, we introduce a graph neural network (GNN) based model, MDNet, to predict the evolution of coordinates and momentum with large time steps. In addition, MDNet can easily scale to a larger system, due to its linear complexity with respect to the system size. We demonstrate the performance of MDNet on a 4000-atom system with large time steps, and show that MDNet can predict good equilibrium and transport properties, well aligned with standard MD simulations.
\end{abstract}

\section{Introduction}


Molecular dynamics (MD) is a simulation technique for analyzing the movement of many-body systems by solving Newton's equation of motion. The resulting trajectories provide a view of dynamic evolution of atoms and molecules, and can be used to analyze the equilibrium and transport properties of the system \cite{frenkel2001understanding}. MD plays a key role in many scientific fields, including materials science~\cite{diddens2013lithium,jabbari2017thermal}, biochemistry~\cite{padhi2021accelerating}, drug discovery~\cite{liu2018molecular}, etc. 

The core of an MD algorithm is to solve Newton's equation of motion at discrete time steps. One of the most frequently used algorithm is the ordinary Verlet integrator~\cite{verlet1967computer}, which predicts the coordinate of atom $i$ at time $t$, $\mathbf{q}_i(t)$, after time step $\Delta t$ by:
\begin{equation}
    \mathbf{q}_i(t+\Delta t) = 2\mathbf{q}_i(t) - \mathbf{q}_i(t-\Delta t) - \frac{\partial \mathcal{U}}{\partial \mathbf{q}_i(t)} \frac{\Delta t^2}{m_i} + \mathcal{O}(\Delta t^4),
\end{equation}
where the potential energy $\mathcal{U}$ of the system is a function of the coordinates $\mathbf{q}$. The time step $\Delta t$ must be set small enough to avoid the contribution of $\mathcal{O}(\Delta t^4)$. In practice, $\Delta t$ is usually chosen at femtosecond scale (1 fs = $10^{-15}$s). Recently, efforts have been made to solve the classical mechanics of many-body systems via machine learning methods \cite{tsai2020learning, kadupitiya2020simulating, gupta2020mind, greydanus2019hamiltonian, chen2019symplectic, chen2021data}. However, all these methods are limited to small systems containing only dozens of atoms, whereas practical MD simulations usually involve thousands to millions of atoms. Therefore, a scalable machine learning approach with linear complexity is highly desired.

In this paper, we propose MDNet, a scalable graph neural network (GNN) model to predict the dynamics of many-body systems. The key contribution is to predict the movement of an atom based on its local environment encoded by a series of descriptors similar as DeePMD~\cite{zhang2018deep, zhang2018end}. In addition,  we introduce momenta in the local environment descriptor and make the model comply with the law of conservation of momenta. We show that MDNet can simulate a system containing thousands of atoms with a time step 128 times larger than that of the Verlet integrator. The resulting equilibrium and transport properties are shown to be consistent with the simulations using the Verlet integrator.

\subsection{Related Work}


The trajectories of MD simulations can be viewed as sequence data hence could be learned by sequence models, such as recurrent neural networks (RNNs). \citeauthor{tsai2020learning} showed that one-dimensional stochastic trajectories generated from higher-dimensional dynamics can be learned by long short-term memory network (LSTM). \citeauthor{kadupitiya2020simulating} demonstrated that LSTM was able to predict MD trajectory for systems containing 16 atoms. Another type of models including HNN \cite{greydanus2019hamiltonian}, SRNN \cite{chen2019symplectic}, GFNN \cite{chen2021data} and others \cite{lutter2019deep, toth2019hamiltonian, zhong2019symplectic, wu2020structure, westermayr2019machine}, manged to incorporate physical knowledge and solve mechanical problems by learning Hamiltonian or other physical quantities. However, all the aforementioned models used the information of all atoms to predict the dynamics, which limits their scalability.

In traditional MD simulations, a common assumption is that atoms only interact with their neighbors. Motivated by this assumption,~\citeauthor{satorras2021n} proposed E(n) equivariant graph neural networks (EGNNs), which model the system by a graph and achieve linear complexity~\cite{satorras2021n}. Compared to EGNN, MDNet use more sophisticated descriptors of local environments and achieve better performance.

\section{Background}

We briefly introduce some basic notations in Hamiltonian mechanics that MDNet is built upon. The Hamiltonian for an N-particle system in $d$ dimensions can be written as:
\begin{equation}
  \mathcal{H}(\mathbf{p}, \mathbf{q}) = \mathcal{T}(\mathbf{p}) + \mathcal{U}(\mathbf{q}),\ \  \mathbf{p}, \mathbf{q} \in \mathbb{R}^{N \times d},
\end{equation}
where $\mathbf{p}, \mathbf{q}$ are the momenta and coordinates of the atoms. The momentum is defined as $\mathbf{p}_i = m_i \mathbf{v_i} $, where $m_i$ and $\mathbf{v}$ are the mass and velocity of atom $i$, respectively. $\mathcal{T}(\mathbf{p}) = \sum_{i=1}^{N} \frac{\mathbf{p}_i}{2m_i}$ is the kinetic energy of the system and $\mathcal{U}(\mathbf{q})$ is the potential energy induced by interaction between atoms. The motion of atoms is determined by Hamilton's equation:
\begin{align}
  \mathbf{\dot{q}_i} = \frac{\partial{\mathcal{H}}}{\partial{\mathbf{p}_i}} = \frac{\mathbf{p}_i}{m_i}, \ \ 
  \mathbf{\dot{p}_i} = -\frac{\partial{\mathcal{H}}}{\partial{\mathbf{q}_i}} = -\frac{\partial{\mathcal{U}}}{\partial{\mathbf{q}_i}} = \mathbf{F}_i. \label{eq:hamilton}
\end{align}
which is a generalization of Newton's equation of motion. The equation is then solved at discrete time steps via numeric integrators, where the velocity Verlet (appendix~\ref{sec:appendix_a})~\cite{swope1982computer} is the most widely used one.



\section{Model Architecture}

\begin{figure}
    \centering
    \includegraphics[width=0.85\textwidth]{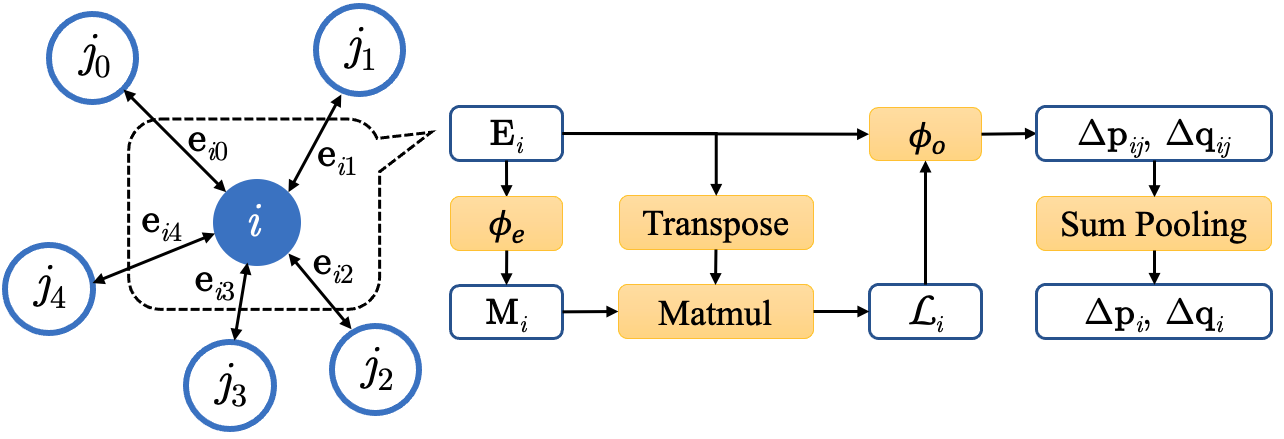}
    \caption{Illustration of model architecture of MDNet.}
    \label{fig:model}
\end{figure}

In Hamilton's equation~\eqref{eq:hamilton}, the variation of $\mathbf{q}$ only depends on $\mathbf{p}$ and vice versa. However, with a larger time step,  atoms interact with each other, thus, $\Delta \mathbf{q}$ depends not only on $\mathbf{p}$, but also $\mathbf{q}$ (and so is $\Delta \mathbf{p}$).
In MDNet, we model $\Delta \mathbf{q}$ and $\Delta \mathbf{p}$ via neural networks:
\begin{align}
  \Delta \mathbf{q}, \Delta \mathbf{p} = \mathrm{MDNet}\left[\mathbf{q}(t), \mathbf{p}(t), \Delta t\right],
\end{align}
and the integration algorithm becomes:
\begin{align}
    \mathbf{p}(t + \Delta t) = \mathbf{p}(t) + \Delta \mathbf{p}, \  \mathbf{q}(t + \Delta t) = \mathbf{q}(t) + \Delta \mathbf{q}.
\end{align}
In MDNet, the system is represented by a graph where each atom is represented by a node. Two nodes are connected by an edge if the distance between them is less than a cut-off distance, $r_c$. 
For an edge between node $i$ and $j$, its edge feature and message is calculated as:
\begin{align}
  \mathbf{e}_{ij} = [\frac{1}{r_{ij}}, \frac{\mathbf{q}_i-\mathbf{q}_j}{r_{ij}^2}, \mathbf{p}_i-\mathbf{p}_j]^T, \ \ \mathbf{m}_{ij} = \phi_e(\mathbf{e}_{ij})
\end{align}
where $r_{ij} = || \mathbf{q}_i-\mathbf{q}_j ||_2$ is the Euclidean distance between node $i$ and $j$, $\mathbf{e}_{ij} \in \mathbb{R}^{m_1}$ and $\mathbf{m}_{ij} \in \mathbb{R}^{m_2}$ . In three-dimensional space, $m_1$ = 7. The features $\frac{1}{r_{i, j}}$ and $\frac{\mathbf{q}_i-\mathbf{q}_j}{r_{ij}^2}$ are motivated from DeePMD~\cite{zhang2018deep,zhang2018end} and $\mathbf{p}_i-\mathbf{p}_j$ brings the information of momenta. Relative coordinates and momenta are used to preserve translation invariance. The edge operator $\phi_e$ is an MLP.  Then, edge features and messages are gathered to nodes. For node $i$ with neighbors $\{j_k\}_{k=0}^n $, the local embedding $\mathcal{L}_i$ is computed as
\begin{align}
  \mathbf{E}_i &= [\mathbf{e}_{i,j_0}, ..., \mathbf{e}_{i, j_{n-1}}], \ \ \mathbf{M}_i = [\mathbf{m}_{i, j_0}, ..., \mathbf{m}_{i, j_{n-1}}], \mathcal{L}_i = \mathbf{M}_i \mathbf{E}_i^T, \ \ \mathcal{L}_i \in \mathbb{R}^{m_2 \times m_1}.
\end{align}
Here permutation invariance is preserved by the multiplication between $\mathbf{E}_i$ and $\mathbf{M}_i$. Finally, the variation of coordinates and momenta are predicted by aggregating the contribution of each edge,
\begin{gather}
  \Delta \mathbf{q}_{ij}, \Delta \mathbf{p}_{ij} = \phi_o(\mathcal{L}_i, \mathbf{e}_{ij}), \\
  \Delta \mathbf{p}_i = \sum_{j \in \mathcal{N}(i)} \Delta {\mathbf{p}}_{ij}, \ \ 
  \Delta \mathbf{q}_i = \sum_{j \in \mathcal{N}(i)} \Delta {\mathbf{q}}_{ij} + \frac{\mathbf{p}_i}{m} \Delta t.
\end{gather}
To ensure the conservation of momentum, for any edge, we randomly choose one direction, e.g. i -> j, compute $\Delta \mathbf{p}_{ij}$ and let $\Delta \mathbf{p}_{ji} = -\Delta \mathbf{p}_{ji}$. This treatment is also applied to the calculation of $\Delta \mathbf{q}_{ij}$'s.


\section{Experiments} \label{Experiments}


We test the performance of MDNet on a liquid argon system containing 4000 atoms. The interatomic interaction is modeled by Lennard-Jones potential~\cite{jones1924determination}:
\begin{align}
  V(r) = 4 \epsilon \left[ (\frac{\sigma}{r})^{12} - (\frac{\sigma}{r})^6 \right]
\end{align}
where r is the distance between two atoms, $\epsilon$ and $\sigma$ are set to 0.2374 kcal/mol and 3.432 \AA\ respectively.
Although there is only one kind of interaction in the system, the dynamics could be rather complicated, with increasing space and time scale. 


{\bf Dataset:} To create the dataset, we run classical MD simulations with LAMMPS using Verlet integrator with a time step of 1 fs. We independently sample 20 trajectories from different initialized configurations, where 15 are used for training, 3 are used for testing and 2 are used for validation.  For each trajectory, we randomly initialize 4000 atoms in a simulation box with periodic boundary condition in all directions. 
The system is first relaxed under 80 K and 1 bar in the NPT ensemble with Nos\'{e}-Hoover (NH) thermostat~\cite{hoover1996kinetic} and barostat for 20000 steps. 
Then, the system is further relaxed under 80 K in the NVT ensemble with NH thermostat for 20000 steps. 
Finally, a NVE simulation is performed for 2560 steps and only the trajectory of this final stage is used. 
In each trajectory, we sample 10 frames $\{\mathbf{q}(t_i), \mathbf{p}(t_i)\}_{i=1}^{10}$ of the system and use $\{\mathbf{q}(t_i + \Delta t), \mathbf{p}(t_i + \Delta t)\}_{i=1}^{10}$ as targets.
According to the time reversibility of MD, the reversed data are also added to the dataset.
Besides, an extra 40960-steps MD simulation in the NVT ensemble is performed to validate roll-out performance.

{\bf Implementation details:} The cutoff radius $r_c$ for constructing the nearest neighbor graph is 7.0 \AA. The MLPs $\phi_e$ and $\phi_o$ both have 4 layers and use ReLU~\cite{nair2010rectified} as activation function. $\phi_e$ has 40 neurons in each hidden layer and $\phi_o$ has 200 neurons in each hidden layer.
Training is carried out for 1000 epoch, batch size 1 with Adam optimizer~\cite{kingma2014adam}. Learning rate starts from $5 \times 10^{-3}$ and decays exponentially at a rate of 0.8. 
The loss function is the mean squared error (MSE) of coordinates and momenta, normalized by their standard deviations respectively.
We compare MDNet to EGNN \cite{satorras2021n} composed of 4 EGCL layers with 64 neurons per layer and Swish activation function~\cite{ramachandran2017searching}. Hyperparameters of EGNN are consist with the original paper.

{\bf One-step results:} We present the one-step root mean squared errors (RMSE) obtained by the Verlet integrator, EGNN and MDNet on validation set in Table \ref{tab:one-step}. 
It is shown that MDNet reports comparable RMSE with Verlet with a small time step (16 fs), and significantly outperforms Verlet with a large time step. As the time step increases, the RMSE of Verlet increases faster than MDNet, due to the contribution of the fourth order error. The performance of EGNN is poor with all the time steps.


\begin{table}[ht]
    \centering
    \begin{tabular}{cccc}
        \toprule
        $\Delta t$ [fs] & Verlet & EGNN & MDNet  \\
        \midrule
        16 & $8.44 \times 10^{-5}, \bf{8.42 \times 10^{-7}}$ & $2.51 \times 10^{-4}, 3.13 \times 10^{-5}$ & $\bf{1.80 \times 10^{-5}}$, $2.32 \times 10^{-6}$ \\
        32 & $6.68 \times 10^{-4}, 6.58 \times 10^{-6}$ & $9.94 \times 10^{-4}, 6.14 \times 10^{-5}$ & $\bf{5.90 \times 10^{-5}}, \bf{4.04 \times 10^{-6}}$ \\
        64 & $5.24 \times 10^{-3}, 5.74 \times 10^{-5}$ & $3.82 \times 10^{-3}, 1.14 \times 10^{-4}$ & $\bf{3.05 \times 10^{-4}}, \bf{1.22 \times 10^{-5}}$ \\
        128 & $3.88 \times 10^{-2}, 1.75 \times 10^{-3}$ & $1.34 \times 10^{-2}, 1.86 \times 10^{-4}$ & $\bf{2.29 \times 10^{-3}}, \bf{5.66 \times 10^{-5}}$ \\
        \bottomrule
    \end{tabular}
    \caption{Validation RMSE for prediction of coordinates in \AA\ (left) and  velocity and \AA/fs (right), respectively. Results obtained from Verlet, EGNN and MDNet with different $\Delta t$.}
    \label{tab:one-step}
\end{table}
\begin{figure}
    \centering
    \includegraphics[width=0.45\textwidth]{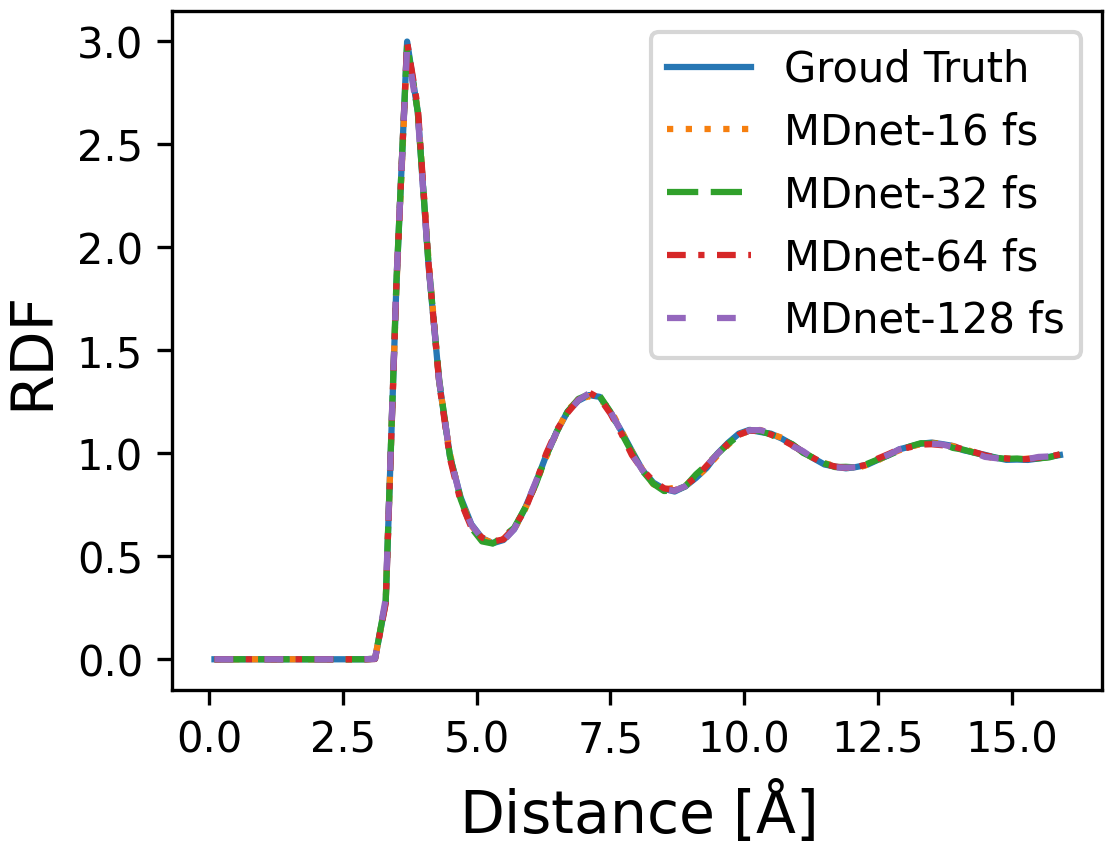}
    \hspace{0.2in}
    \includegraphics[width=0.45\textwidth]{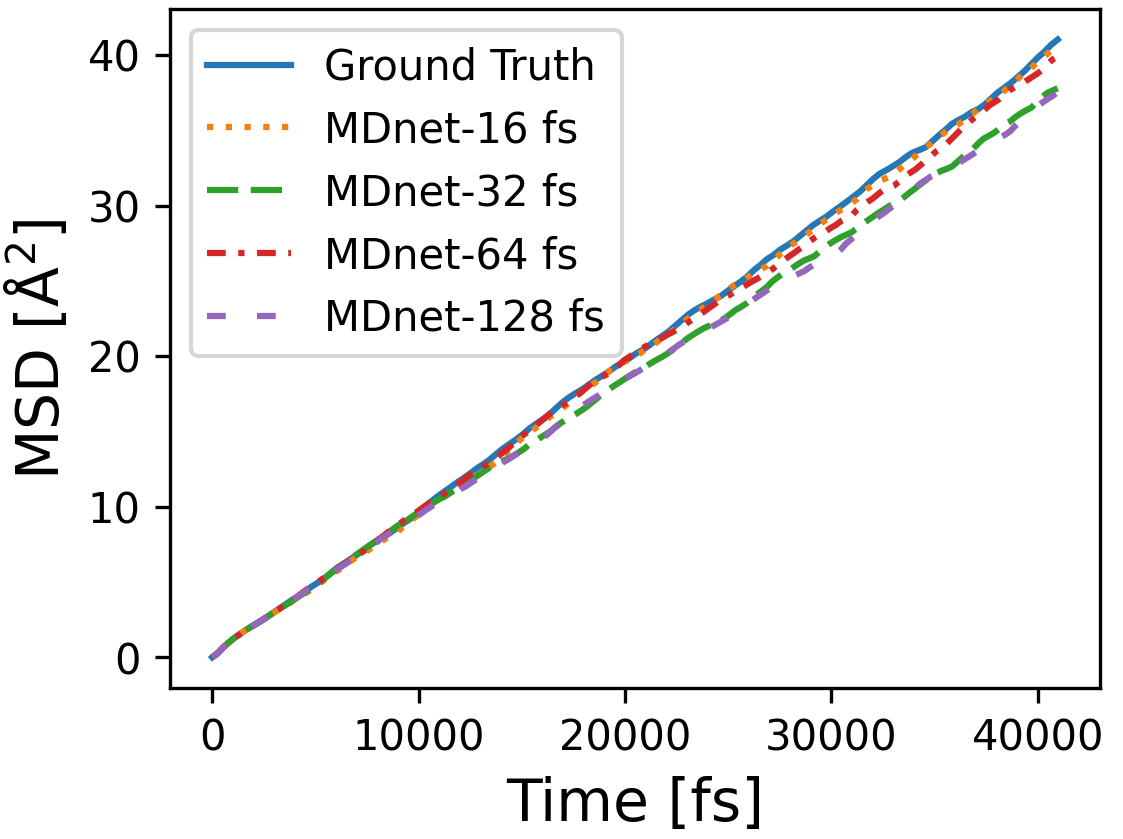}
    \caption{RDFs and MSDs computed from ground truth and MDNet with different $\Delta t$.}
    \label{fig:results}
\end{figure}

{\bf Roll-out performance:} MD simulations can be performed by adopting one-step prediction iteratively, which is known as roll-out. 
We use two metrics to evaluate the roll-out performance of MDNet: radial distribution function (RDF) \cite{levine2011fast} and mean-squared displacement (MSD) \cite{afandak2017ion, wang2020lithium}. 
where RDFs provide good descriptions of equilibrium structure and MSD is a measure of transport properties of simulation systems.
To compute RDFs, we conduct 40000 fs simulation and take 8 frames from the subsequent trajectory with a gap of 128 fs. The final RDFs are averaged over these 8 frames. 
As shown in Figure \ref{fig:results} (left), RDFs obtained from MDNet with different time steps coincide with ground truth, which show correct equilibrium structures. 
To compute MSD, the trajectory is saved with a resolution of 256 fs and calculate displacement against the first frame.
As shown in Figure \ref{fig:results} (right), MSDs obtained from MDNet are linear with time and are in good agreement with ground truth, despite that the slopes with large time steps are slightly smaller. In summary, MDNet can reproduce the equilibrium structures and transport properties of the simulation system with large time steps. We also compare the time cost of roll-out and show that MDNet enjoys faster roll-out speed compared to Verlet, thanks to the large time step. The comparison of time cost is in the appendix~\ref{sec:appendix_b}.

\section{Conclusion and Discussion}

We introduce MDNet, a GNN model that can predict the dynamics of many body system with large time steps. 
We demonstrate that MDNet exhibites good precision on one-step prediction with a large time step, and is able to generate MD trajectories. The generated trajectories are characterized by RDF and MSD, which showed good agreement with the ground truth. 
It is worth mentioning that MDNet only takes local environments as feature, so it could be generalized to larger systems with the same species of atoms without further training and we plan to investigate this in our future work. 

There are several future directions. Firstly, the total energy is not guaranteed to be conserved during roll-out of MDNet. Incorporating thermostat into MDNet is necessary to control the temperature and total energy. Secondly, The Lennard-Jones system presented in this paper is the simplest case in MD applications. With the introduction of chemical bonds, the system will exhibit multiple characteristic times and the dynamics will be much more difficult to predict. The extension of MDNet to complicated systems is worthy of further study. 





\bibliography{MLMD.bib}

\clearpage

\appendix

\section{The Velocity Verlet Algorithm}
\label{sec:appendix_a}
The algorithm of the most commonly used integrator, velocity Verlet, is:
\begin{align}
  \mathbf{p}_i(t+\frac{1}{2}\Delta t) &= \mathbf{p}_i(t) + \frac{1}{2}\mathbf{F}_i(t)\Delta t,\nonumber\\
  \mathbf{q}_i(t + \Delta t) &= \mathbf{q}_i(t)+\frac{\mathbf{p}_i(t+\frac{1}{2}\Delta t)}{m_i}\Delta t,\\
  \mathbf{p}_i(t+\Delta t) &= \mathbf{p}_i(t+\frac{1}{2}\Delta t)+\frac{1}{2}\mathbf{F}_i(t + \Delta t)\Delta t. \nonumber
\end{align}
In practice, $\mathbf{p}_i(t+\Delta t)$ may not need to be calculated explicitly. Then, the coordinates and momenta are updated alternately, which is also known as leapfrog integration.

\section{Comparison of Time Cost}
\label{sec:appendix_b}

We demonstrate the time cost of roll-out of MDNet and Verlet algorithm here. 

\begin{table}[h]
    \centering
    \begin{tabular}{ccc}
        \toprule
         & forward time [s] & time cost per fs [s]  \\
        \midrule
        Verlet-1 fs & $3.82 \times 10^{-3}$ & $3.82 \times 10^{-3}$\\
        MDNet-16 fs & $4.01 \times 10^{-2}$ & $2.51 \times 10^{-3}$ \\
        MDNet-128 fs & $4.03 \times 10^{-2}$ & $3.15 \times 10^{-4}$ \\
        \bottomrule
    \end{tabular}
    \caption{The time cost of Verlet and MDNet for one-step prediction and 1 fs simulation. For fair comparision, Verlet algorithm is implemented by Tensorflow~\cite{tensorflow2015-whitepaper} and MDnet by DGL~\cite{wang2019dgl} with Tensorflow backend. The computation time is evluated on a single Tesla P4.}
    \label{tab:cost}
\end{table}

\end{document}